\documentclass[submission]{eptcs}

\providecommand{\event}{appFM~2021}
\usepackage{breakurl}
\usepackage{underscore}

\newcommand{\question}[1]{\noindent$\alpha$FM: \emph{#1}\nopagebreak[3]\medskip}

\title{Is CADP an Applicable Formal Method?}
\author{\mbox{Hubert Garavel \hspace{2em} Fr\'ed\'eric Lang \hspace{2em} Radu Mateescu \hspace{2em} Wendelin Serwe}
\institute{Univ. Grenoble Alpes, Inria, CNRS, Grenoble INP\footnote{Institute of Engineering Univ. Grenoble Alpes}, LIG, 38000 Grenoble, France}
\email{\{Hubert.Garavel,Frederic.Lang,Radu.Mateescu,Wendelin.Serwe\}@inria.fr}
}
\def\titlerunning{Is CADP an Applicable Formal Method?}
\def\authorrunning{H. Garavel, F. Lang, R. Mateescu, W. Serwe}

\begin{document}

\maketitle

\begin{abstract}
  CADP is a comprehensive toolbox implementing results of concurrency theory.
  This paper addresses the question, whether CADP qualifies as an applicable formal method, based on the experience of the authors and feedback reported by users.
\end{abstract}

\section{Introduction}
\label{sec:introduction}

Formal Methods~\cite{Garavel-Graf-13,Garavel-terBeek-vandePol-20} are a wide spectrum of languages, techniques, and tools, strongly rooted in theory, for the design, analysis, and validation of systems.
CADP (Construction and Analysis of Distributed Processes)\footnote{\url{http://cadp.inria.fr}}~\cite{Garavel-Lang-Mateescu-Serwe-13} is a comprehensive toolbox that implements the results of concurrency theory.
Started in the mid 80's, CADP initially consisted of only two tools.
Over the past 35 years, CADP has been continuously improved and extended, and contains now more than 50 tools and almost 20 software components.
CADP offers functionalities covering the entire design cycle of asynchronous concurrent systems: specification, interactive simulation, rapid prototyping, verification, testing, and performance evaluation.
For verification, CADP supports the three essential approaches existing in the field: model checking, equivalence checking, and visual checking.
To deal with complex systems, CADP implements a wide range of verification techniques (reachability analysis, static analysis, on-the-fly, compositional, and distributed verification) and provides a scripting language for describing elaborate verification scenarios.
In addition, CADP supports many different specification languages.

CADP benefits from a worldwide user community: The CADP web site lists more than 200 published case studies in numerous application domains and more than 100 published third-party tools.
Most of them were not carried out by the authors of the toolbox.
Some case studies were conducted with only a minimal amount of training and/or consulting, by almost novice users of formal methods~\cite{Bouzafour-Renaudin-Garavel-et-al-18}.
Many applications of CADP were carried out by industrial companies, including Airbus, Bull, Crouzet/Innovista Sensors, Google, Nokia Bell Labs, Orange Labs, STMicroelectronics, and Tiempo Secure.

The rest of this paper is organized as follows: Sections~\ref{sec:applicability} to~\ref{sec:evaluation} address the questions listed in the topics of interest\footnote{\url{https://sites.google.com/view/appfm21/scope}} of the \event{} workshop.
The answers to each question are not only based on our own experience in teaching as well as academic and industrial collaborative projects, but also on feedback reported in publications on case studies and third-party tools connected to CADP.
Section~\ref{sec:conclusion} concludes the discussion with additional criteria considered important by the authors.

\section{Applicability}
\label{sec:applicability}

\question{How can the approach be applied in practice?}

\noindent
CADP is used by mainly two audiences.
The first one consists of students learning concurrency theory concepts, for which CADP provides concrete representations (concurrent processes, synchronization and communication, temporal logic formulas, etc.).
For example, the abstract notion of automaton is instantiated as files in the BCG (Binary Coded Graphs) format, which can be visualized, minimized modulo bisimulation relations, modified by hiding or renaming transition labels, etc.

The second audience consists of scientists or industry engineers, who build complex and critical systems, and strive to follow a rigorous design process and to obtain guarantees of correct functioning and performance.
Here, CADP assists in all the main design phases: specification (modeling concurrent processes and their interactions), design-time analysis (verification and performance evaluation), validation of an existing implementation (co-simulation, conformance test generation).
Design-time analysis of a formal model has the advantage of early error detection, which significantly reduces the cost of errors.

Among the case studies listed on the CADP web site, the tools are most frequently used for the formal specification and modeling of a system, which are then verified by model checking temporal logic properties and/or equivalence checking against a reference model (e.g., the expected service of a protocol).
Some case studies additionally take advantage of more specific tools, e.g., for
performance evaluation~\cite{Garavel-Hermanns-02,Boede-Herbstritt-Hermanns-et-al-06,Crouzen-vandePol-Rensink-08,Chehaibar-Zidouni-Mateescu-09,Coste-Hermanns-Lantreibecq-Serwe-09,Foroutan-Thonnart-Hersemeule-Jerraya-10,Mateescu-Serwe-13,Wu-Yang-Katoen-16},
(conformance) test generation~\cite{Kahlouche-Viho-Zendri-99,DuBousquet-Ramangalahy-et-al-00,Garavel-Viho-Zendri-01,Scollo-Zecchini-05,Turner-05a,Chimisliu-Wotawa-13-b,GrafBrill-Hermanns-Garavel-14,Lantreibecq-Serwe-14,Kriouile-Serwe-15,Bozic-Marsso-Mateescu-Wotawa-18},
or generation of an executable prototype from the formal model~\cite{DuBousquet-Ramangalahy-et-al-00,Garavel-Viho-Zendri-01,Su-Bowman-Barnard-Wyble-08,Garavel-Thivolle-09,Lantreibecq-Serwe-14,Serwe-15,Garavel-Serwe-17}.

\section{Automation}
\label{sec:automation}

\question{Which tool support is proposed? If abstraction is needed, how is it automated?}

\noindent
Tool support has always been central in CADP, which is a toolbox rather than a monolithic tool supporting a single language and methodology (e.g., the B-method).
CADP has completely automated tools for simulation, for which abstraction is not desirable, because an accurate model is needed.

Concerning enumerative verification, although often advertised as a push-button technique, practice shows that, like for theorem proving, a fair amount of human intervention may be needed (to fight state space explosion).
For now, not everything is automatable, nor will it be in a foreseeable future.
One still relies on experts, on their knowledge and competence to select an appropriate verification strategy.
In general, on-the-fly techniques are used to quickly find bugs in initial versions of a model,
whereas more involved strategies may be better to generate an abstract model (reduced for an appropriate bisimulation), on which temporal logic properties are to be verified formally.
The most prominent technique implemented in CADP is compositional verification, which is a divide-and-conquer approach applying compositions and abstractions incrementally.
CADP offers a collection of known abstractions, which are generic recipes that succeed in many cases.
But, for some particularly involved problems, one must design specific abstractions after a careful analysis of the problem~\cite{Garavel-Lang-Mateescu-15,Lang-Mateescu-Mazzanti-19,Lang-Mateescu-Mazzanti-20}.
Compositional verification is supported by SVL (Script Verification Language)~\cite{Garavel-Lang-01}, providing, among others, also an automatic heuristic, called smart reduction~\cite{Crouzen-Lang-11}, which selects an order for compositions.
In the context of semi-composition~\cite{Krimm-Mounier-97}, the EXP.OPEN tool~\cite{Lang-05} can automatically compute behavioral interfaces~\cite{Lang-06} and check the correctness of manually provided ones.
Taking advantage of the integration of shell commands in SVL, CEGAR (Counter-Example Guided Abstraction Refinement)-style loops have been used to automatically reduce the model~\cite{Kriouile-Serwe-15}.

Turning informal specifications into formal models and/or properties is hardly automatable.
However, the modeling languages of CADP ease this step by supporting concurrent programs with complex and/or dynamic data structures (records, unions, lists, trees, etc.).
These languages are also convenient targets to compile domain-specific modeling languages~\cite[Figure~4]{Garavel-Lang-Mateescu-Serwe-13}: dedicated compilers have been developed for instance for
FSP~\cite{Salaun-Kramer-Lang-Magee-07},
CHP~\cite{Garavel-Salaun-Serwe-09},
RT-UML~\cite{SaquiSannes-Apvrille-09},
the (applied) $\pi$-calculus~\cite{Mateescu-Salaun-13},
dynamic fault trees~\cite{Guck-Spel-Stoelinga-15}, or
AADL~\cite{Mkaouar-Zalila-Hugues-Jmaiel-20}.
It is worth noting that the asynchronous subset of SystemVerilog can be translated almost line-by-line into LNT~\cite[Fig.~2]{Bouzafour-Renaudin-Garavel-et-al-18}.

In some cases, the rich data domains must be reduced by abstraction.
Although not automatic, this is facilitated by the possibility to redefine basic data types~\cite{Serwe-15} or adding constraints~\cite[Section~V.C.1]{Bouzafour-Renaudin-Garavel-et-al-18}.

\section{Integration}
\label{sec:integration}

\question{If several approaches are to be applied in an integrated way; if the approach is to be used with a modeling technique or programming language; if the approach is to be integrated into an engineering process, what are the benefits?}

\noindent
CADP is modular, providing many tools, libraries, formats, and languages.
The formats and languages cover different abstraction levels (from process calculi to explicit automata descriptions for models, and from MCL formulas to XTL programs for properties).
The tools and libraries provide well-defined functionalities, which can be combined and interconnected in various ways to improve the overall analysis approach by fighting state space explosion.
The OPEN/C\AE{}SAR architecture~\cite{Garavel-98} separates language-dependent and language-independent aspects and has been identified as a key for smooth integration (``\emph{the OPEN/C\AE{}SAR interface has been underlying the success of CADP}''~\cite{Blom-vandePol-Weber-10}).

CADP tools can be combined in several ways, and at various levels, e.g., through verification scripts, translations between languages, and tool interconnection using application programming interfaces, e.g., OPEN/C\AE{}SAR, BCG, etc.
Data-handling C~code (e.g., existing optimized implementations of complex data structures, such as the directories of a cache-coherence protocol) can be called from the models.
CADP provides comprehensive documentation for all tools and libraries, in the form of manual pages, totaling more than 800 pages.
The benefit of this open, documented architecture is witnessed by the numerous third-party tools, which, taking advantage of functionalities offered by CADP, provide domain specific tools or particular verification features.
This architecture also simplifies the implementation of new prototypes~\cite{Lang-Mateescu-13,Marsso-Mateescu-Serwe-18}.

\section{Scalability}
\label{sec:scalability}

\question{How can the approach be applied at scale, for example, using composition and refinement?}

\noindent
The main limiting factor for enumerative verification is the amount of available memory.
Thus, the CADP tools and libraries are optimized to reduce memory usage before execution time.
CADP is implemented mostly using shell scripts and the C programming language, which enables a fine control on the memory usage, enabling low-level optimization to reduce the memory footprint of each state at bit-level.
The most prominent example is the BCG format, introduced in 1994, which enables compact storage of automata with up to $10^{13}$ states and transitions.

To benefit from the combined memory available in clusters and grids, CADP provides distributed tools for state space generation~\cite{Garavel-Mateescu-Bergamini-et-al-06,Garavel-Mateescu-Serwe-12} (possibly combined with on-the-fly reductions) and resolution of Boolean equation systems~\cite{Joubert-Mateescu-04}, into which a wide range of verification problems can be encoded.

However, compositional techniques~\cite{Garavel-Lang-Mateescu-15}, the most prominent among which is compositional state space construction, are the major asset of CADP for scalability.
``\emph{The advantage of using compositional construction in terms of space and time is apparent. Stepwise minimization keeps the size of state spaces low. This, in turns, reduces the duration of the minimization time in the next step, and so on, thus saving significant amount of time.}''~\cite[Section~VI.C]{Boede-Herbstritt-Hermanns-et-al-06}
Compositional techniques rely on the compositional properties of process calculi and the fact that interesting equivalences are congruences for the principal composition operators.
CADP also relies on adequacy results between automata equivalences and temporal logics.
Besides compositional construction, model checking can also be applied compositionally.
This technique, called partial model checking, has been implemented as a companion tool to CADP~\cite{Lang-Mateescu-13}.
Having access to many verification strategies allowed to win gold medals in the parallel tracks of the 2019 and 2020 editions of the RERS challenge (Rigorous Examination of Reactive Systems).\footnote{\url{http://rers-challenge.org}}

In a comparison of formal verification tools for a railway problem~\cite{Mazzanti-Ferrari-18}, CADP ranked among both the fastest tools and among those with the lowest memory requirements.

\section{Transfer}
\label{sec:transfer}

\question{How is teaching or training to be organized to transfer the approach?}

\noindent
One of the most considered research priorities for formal methods research is the applicability and acceptability of tools~\cite{Garavel-terBeek-vandePol-20}.
Indeed, formal methods are reputed to have a steep learning curve hindering their easy acceptance outside academia.
Since many years, the input languages of CADP are continuously being improved to flatten the learning curve, so as to enable our industrial partners to model their confidential systems in-house with as little guidance as possible.

Since 2010, the recommended modeling language of CADP is LNT~\cite{Champelovier-Clerc-Garavel-et-al-10-v7.0}, which is as close as possible to languages known by engineers and students~\cite{Garavel-Lang-Serwe-17}, using a familiar syntax rather than algebraic notations, enabling a novice user to build upon known notions and to focus on important concepts.

For temporal logics, CADP provides MCL (Model Checking Language)~\cite{Mateescu-Thivolle-08}, which takes advantage of regular expressions and offers the possibility of macro definitions.
Together with a rich set of macro libraries, MCL hides most complex constructions (in particular fixed-point operators) from the user.

\section{Usefulness}
\label{sec:usefulness}

\question{How is usefulness achieved?  Is the approach effective?  What would have been different if a conventional or non-formal alternative was used (e.g., relative fault-avoidance or fault-detection effectiveness)?}

\noindent
Usefulness of CADP is witnessed by the more than 200 case studies and 100 research tools published.
Some case studies explicitly confirm that formal analyses are effective in the sense that they enable errors to be detected earlier:
``\emph{Architects detected a limitation in the IP [...]. This limitation manifests in a subset of the counterexamples for the data integrity property we verified 20 months before.}''~\cite[Section~6.2]{Kriouile-Serwe-15}
``\emph{It was agreed that the use of formal methods clearly increased the quality of the design, by detecting errors and by showing that, after correction, these errors would disappear from the modified versions of the cache coherency protocol.}''~\cite{Garavel-Viho-Zendri-01}

Effectiveness can even be improved by leveraging the modeling effort over several activities, e.g., formal verification, performance evaluation, and conformance testing.
For the latter, it yields better coverage and uncovers bugs: ``\emph{We should notice that our 306 extracted tests trigger those checks 16 times, whereas the other tests of the [...] test library never trigger these checks}'' and ``\emph{we observe that the coverage of the verification plan increased significantly and that the coverage of the svunit part of the verification plan is complete (100\%), i.e., all the aspects corresponding to system-level behaviors are tested.}''~\cite[Section~6.2]{Kriouile-Serwe-15}
``\emph{The time spent in specifying the Bull's CC\_NUMA architecture, formalizing test purposes and generating the test cases with TGV is completely paid by the better correctness and the confidence to put in the implementation. This approach permitted to detect 5 bugs.}''~\cite{Kahlouche-Viho-Zendri-99}

The capability to generate counterexamples and witnesses, common to all enumerative techniques, is also judged an interesting asset.
``\emph{Another important advantage of using CADP is that, when a property does not hold, the model checking algorithm generates a counter-example, i.e., an execution trace leading to a state in which the property is violated. This ability to generate counterexamples can be exploited to pinpoint the cause of an error and possibly correct it.}''~\cite{Martinelli-Mercaldo-Nardone-et-al-19}

\section{Ease of Use}
\label{sec:ease}

\question{How is ease of use achieved?  Is the approach efficient? How was its usability and maturity assessed (e.g., abstraction effort, proof complexity, productivity) and what are the results?}

\noindent
The transfer-facilitating aspects mentioned in Section~\ref{sec:transfer}, in particular concerning the input languages, are also crucial for the ease of use.
Process calculi and concurrency theory are useful, but notoriously difficult to grasp. One goal of CADP was to make a shift from abstract mathematics to concrete computer science, and a significant effort was invested in this direction:

For modeling languages, LOTOS~\cite{ISO-8807} (based on algebra for both processes and data types) was progressively replaced by LNT (combining imperative and functional programming), which is closer to the practical needs and intuitive for users.
$80\%$ of LNT concepts are well-known to programmers, only $20\%$ being related to concurrency (e.g., processes, parallel composition, and rendezvous).
Moreover, functions and processes have a unified syntax, which makes a great step forward from previous languages based on process calculi~\cite{Garavel-Lang-Serwe-17}: ``\emph{Although modeling the DTD in a classical formal specification language, such as LOTOS~\cite{ISO-8807}, is theoretically possible, using LNT made the development of a formal model practically feasible. In particular, features such as predefined array data-types, loops, and modifiable variables helped to obtain a model easily understandable by hardware architects.}''~\cite[Introduction]{Lantreibecq-Serwe-14}.

For tools, besides the command-line syntax documented in manual pages, CADP provides for more novice users a graphical user interface with contextual menus for easy invocation of the various tools: ``\emph{Through the Xeuca interface (see Figure~4) [the] CADP toolbox allows an easy access to the offered functionalities.}''~\cite[Section~5.3]{AmeurBoulifa-Cavalli-Maag-19}.
The SVL language enables a versatile description of complex verification scenarios, applies automatically various heuristics for combating state explosion, and provides support for batch-mode execution of verification experiments.
``\emph{CADP provides a scripting language, SVL~\cite{Garavel-Lang-01}, which is particularly convenient to experiment with different strategies to alternate construction and minimization steps.}''~\cite[Section~V]{Boede-Herbstritt-Hermanns-et-al-06}

In~\cite{Mazzanti-Ferrari-18}, the authors highlight among the advantages of CADP the imperative modeling style of LNT fitting their state-machine oriented representation and the expressive power of the property definition language MCL (that subsumes both LTL, albeit at a non-linear cost, and CTL).
``\emph{CADP [tools] can be used in their default configuration without having to specify any particular evaluation choice.}''~\cite{Mazzanti-Ferrari-18}

\section{Evaluation}
\label{sec:evaluation}

\question{Why will the approach be useful for a wide range of critical applications? }

\noindent
Being based on the generic concepts of concurrency theory, and due to its flexible architecture and open interfaces, CADP has already been used in various application domains, including typical critical systems (for instance, avionics, autonomous vehicles, cryptography, embedded software, hardware design, railways, and security), but also more exotic applications, such as human cognitive processes~\cite{Su-Bowman-Barnard-Wyble-08}.

To further promote the use of formal methods, we make results of some case studies available to the research community, by contributing models in various forms as challenges to competitions (such as the \emph{Model Checking Contest}\footnote{\url{https://mcc.lip6.fr/}}), repositories (such as the \emph{Models for Formal Analysis of Real Systems} repository\footnote{\url{http://www.mars-workshop.org/repository.html}}), or benchmarks.\footnote{\url{http://cadp.inria.fr/resources}}

\section{Conclusion}
\label{sec:conclusion}

CADP is an efficient and usable verification toolbox developed on and for the long run by a small team of researchers, with limited resources.
It was always meant to be a \emph{true} software, intended for real users, rather than a mere research prototype to accompany publications.
This has implications beyond the questions addressed before:

\smallskip\noindent\textbf{Software primacy.} There is a true difference between a scientific publication and a software tool.
Ideally, a publication is a fixed artefact, which is rarely updated after print.
On the other hand, software maintenance represents the largest part of the software life-cycle: without maintenance, the software tool soon deprecates and becomes unusable, as the processors, operating systems, and software libraries continue their evolution.

\smallskip\noindent\textbf{Stability.} Learning the CADP tools is an investment from our users, and we want to preserve such investment over the years.
All the prototype tools we develop are not automatically integrated in CADP.
But once a tool is integrated, this means that its functionality has been found to be useful, and we try to maintain it over the years.
Sometimes, certain tools are eliminated (e.g., ALDEBARAN), but we replace them with equivalent tools and shell scripts that preserve backward compatibility.

\smallskip\noindent\textbf{Testing.} We have accumulated hundreds of thousands of artefacts (programs, models, formulas, automata, etc.) that are routinely used to test the quality and stability of CADP after each modification.

\smallskip\noindent\textbf{Documentation.} Each tool and library is documented by a detailed manual page.
CADP comes with a set of demonstration examples covering various application domains, a collection of frequently asked questions, and a user forum\footnote{\url{http://cadp.forumotion.com}} to create a user community and archive answers to specific questions in an easily accessible way.
Besides easing transfer, these features also make CADP popular for teaching.
There is no textbook, since it would give a frozen vision, as the toolbox is evolving (still keeping compatibility with previous versions).
Instead, all resources are available on the CADP web site.

We conclude with words of CADP users:
``\emph{We exploit CADP since it is a popular toolbox maintained, regularly improved, and used in many industrial projects, as a verification framework.}''~\cite{Martinelli-Mercaldo-Nardone-et-al-19}
The overall usability is so good that ``\emph{main barriers [to a more widespread inclusion] are the limited support for development functionalities, such as traceability, and other process-integration features.}''~\cite{Ferrari-Mazzzanti-Basile-et-al-21}

\end{document}